\documentclass[12pt]{article}
\usepackage{authblk}
\usepackage{array}

\usepackage{amsbsy,amsmath,amsthm,latexsym,amssymb}
\usepackage{graphicx,stmaryrd,sectsty}
\usepackage{setspace}
\usepackage[font=small,labelfont=bf]{caption}
\usepackage{verbatim}
\usepackage{graphicx}
\usepackage{caption}
\usepackage{subcaption}
\usepackage{float,color}
\usepackage{url}
\newcommand{\bc}{\begin{center}}
	\newcommand{\ec}{\end{center}}
\newcommand{\bfr}{\begin{flushright}}
	\newcommand{\efr}{\end{flushright}}

\newcommand{\be}{\begin{enumerate}}
	\newcommand{\ee}{\end{enumerate}}
\newcommand{\bi}{\begin{itemize}}
	\newcommand{\ei}{\end{itemize}}
\newcommand{\bd}{\begin{description}}
	\newcommand{\ed}{\end{description}}
\newcommand{\beq}{\begin{equation}}
\newcommand{\eeq}{\end{equation}}
\newcommand{\bea}{\begin{eqnarray}}
\newcommand{\eea}{\end{eqnarray}}

\newcommand{\bfi}{\begin{figure}}
	\newcommand{\efi}{\end{figure}}
\newcommand{\bay}{\begin{array}{l}}
	\newcommand{\eay}{\end{array}}

\newcommand{\cref}[1]{(\ref{#1})}   

\sectionfont{\normalsize}
\subsectionfont{\normalsize}
\subsubsectionfont{\normalsize}

\begin{document}



\begin{titlepage}
	\clearpage\thispagestyle{empty}
	\noindent
	\hrulefill
	\begin{figure}[h!]
		\centering
		\includegraphics[width=2 in]{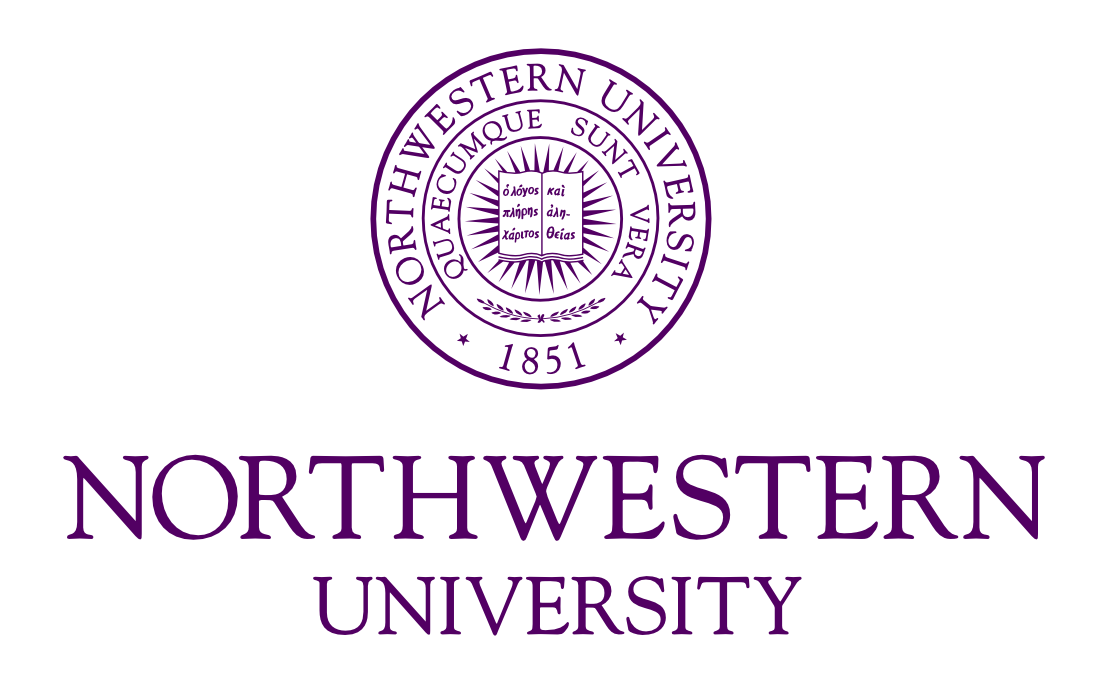}
	\end{figure}
	\begin{center}
		{
			{
				{\bf Center for Sustainable Engineering of Geological and Infrastructure Materials} \\ [0.1in]
				Department of Civil and Environmental Engineering \\ [0.1in]
				McCormick School of Engineering and Applied Science \\ [0.1in]
				Evanston, Illinois 60208, USA
			}
		}
	\end{center} 
	\hrulefill \\ \vskip 2mm
	\vskip 0.5in
	\begin{center}
		{\large {\bf ELASTIC MICROPLANE FORMULATION FOR TRANSVERSELY ISOTROPIC MATERIALS}}\\[0.5in]
		{\large {\sc Congrui Jin, Marco Salviato, Weixin Li, Gianluca Cusatis}}\\[0.75in]
		{\sf \bf SEGIM INTERNAL REPORT No. 16-8/707E}\\[0.75in]
	\end{center}
	\noindent {\footnotesize {{\em Submitted to Journal of Applied Mechanics \hfill August 2016} }}

\newpage
\clearpage \pagestyle{plain} \setcounter{page}{1}









  



\end{titlepage}

\title{Elastic Microplane Formulation for Transversely Isotropic Materials}

\author[1]{\small Congrui Jin}
\author[2]{Marco Salviato\thanks{Corresponding author}}
\author[3]{Weixin Li}
\author[3]{Gianluca Cusatis}

\affil[1]{\footnotesize Department of Mechanical Engineering,\\
	State University of New York at Binghamton,\\ Vestal, NY 13902, USA.}
\affil[2]{\footnotesize William E. Boeing Department of Aeronautics and Astronautics,\\
	University of Washington,\\ Seattle, WA 98195, USA.}	
\affil[3]{\footnotesize Department of Civil and Environmental Engineering,\\
	Northwestern University,\\ Evanston, IL 60208, USA.}	
\date{}
\maketitle	
\let\thefootnote\relax\footnote{\footnotesize \textit{Email addresses:} cjin@binghamton.edu (Congrui Jin), salviato@aa.washington.edu (Marco Salviato), w.li@u.northwestern.edu (Weixin Li), g-cusatis@northwestern.edu (Gianluca Cusatis)
}

	{\small \noindent {\bf   Abstract}: 
	This contribution investigates the extension of the microplane formulation to the description of transversely isotropic materials such as shale rock, foams, unidirectional composites, and ceramics. Two possible approaches are considered: 1) the spectral decomposition of the stiffness tensor to define the microplane constitutive laws in terms of energetically orthogonal eigenstrains and eigenstresses; and 2) the definition of orientation-dependent microplane elastic moduli. It is shown that the first approach provides a rigorous way to tackle anisotropy within the microplane framework whereas the second approach represents an approximation which, however, makes the formulation of nonlinear constitutive equations much simpler. The efficacy of the second approach in modeling the macroscopic elastic behavior is compared to the thermodynamic restrictions of the anisotropic parameters showing that a significant range of elastic properties can be modeled with excellent accuracy. Further, it is shown that it provides a very good approximation of the microplane stresses provided by the first approach, with the advantage of a simpler formulation.
	
	It is concluded that the spectral stiffness decomposition represents the best approach in such cases as for modeling unidirectional composites, in which accurately capturing the elastic behavior is important. The introduction of orientation-dependent microplane elastic moduli provides a simpler framework for the modeling of transversely isotropic materials with remarked inelastic behavior, as in the case, for example, of shale rock.
}

\section{Introduction}

Quasi-brittle materials are defined as those materials that exhibit no or negligible plastic strain prior to failure \cite{0}. The formation and growth of fracture process zone are considered to be responsible for both the softening
behavior observed in the post-peak stress-strain curve
and the development of plastic irreversible strains \cite{0+}. The presence of anisotropy in quasi-brittle materials is very common. For example, the response of rigid foams is usually anisotropic. During the foaming process viscous forces cause the cells to be elongated in the rise direction, and the material response will be therefore stiffer in this direction \cite{14}. A special case of anisotropy is transverse isotropy, which contains a plane of isotropy, implying that the material can be rotated with respect to the loading direction about one axis without measurable effect on the material's response. Due to its high symmetry and relative simplicity in mathematical formulae, transversely isotropic medium has become one of the most studied
anisotropic media in the literature. Fiber-reinforced composites with all fibers being in parallel can be regarded as transversely isotropic, and many sedimentary rocks, such as shales, slates, siltstones, claystones, and mudstones, are best described as transversely isotropic media with the symmetric axes perpendicular to bedding. Such bedding planes affect the strength and deformational behaviors of
the rock with orientation to the applied stresses. 

Elastic transverse isotropy is the subject of the present contribution which investigates the extension of the microplane formulation to this type of anisotropy. Two possible approaches are compared: 1) the spectral decomposition of the stiffness tensor to define the microplane constitutive laws in terms of energetically orthogonal eigenstrains and eigenstresses; and 2) the definition of orientation-dependent microplane elastic moduli.

\section{Background of Microplane Model}

The microplane model describes the material behavior at the mesoscopic scale by
formulating the constitutive laws in terms of
stress and strain vectors acting on individual
microplanes of all possible orientations at a given material point \cite{1,2,4,3,5,6,7,8,9,10,11,12,13,14,15,M7,M7+,kidar,M6,xinwei}, instead of using a traditional tensorial constitutive model. These microplanes may be imagined to represent damage planes or weak planes in the mesoscale structures, such as contact layers between aggregate pieces in concrete or defects in composite laminates. 

The microplane concept has known a long history. The characterization of the material
behavior on different material planes was first
suggested by Mohr \cite{16} in 1900. This
idea was then advanced by Taylor \cite{17}, and applied to develop the slip theory of plasticity by Batdorf and Budiansky \cite{17+}. Later, it was extended by Ba\v zant and his co-workers to model quasi-brittle materials exhibiting softening damage \cite{1,2}. Since then the microplane model for concrete has been studied extensively, and evolved through several progressively improved versions \cite{1,2,5,6,3, 8, 7,9, 10,M6,M7,M7+,kidar}. Numerous advantages of microplane models were reviewed in Brocca and Ba\v zant \cite{bro} and Cusatis et al. \cite{cu1}. The main appealing aspect of this approach is its conceptual simplicity, i.e., once the general algorithm for the relationship between microplane quantities and macroscopic tensors has been established, formulating a constitutive
law is intuitive, since all the quantities involved always have an immediate physical meaning. Oriented phenomena, such as friction and cracking, can be realistically simulated. Besides, the microplane model automatically exhibits the vertex effect, which has not been captured by any usable tensorial models, and the interaction of microplanes accurately captures all the cross effects, such as shear dilatancy and pressure sensitivity. This also allows simulating damage-induced anisotropy quite simply. Despite the fact that the adoption of the microplane modeling approach is usually computationally expensive compared to the classical tensorial models, systems of millions of finite elements have been successfully solved using microplane model for concrete \cite{4}. Microplane models have also been developed for other complex materials such as rock \cite{11}, cemented soils \cite{18}, clay \cite{12}, rigid foam \cite{14}, fiber reinforced concrete \cite{15}, shape memory alloy \cite{sma}, and fiber composites (prepreg laminates \cite{cu1,cu2} and braided composites \cite{br}). Finally, it is worth noting that the constitutive relations prescribed
on the microplanes, which are lumped into a single material
point in the microplane models, can also be used in an explicitly
mesoscale model on planes of various orientations separating the
neighboring aggregates embedded in a cement mortar matrix, as it has been done in the recently developed Lattice Discrete Particle Model (LDPM) \cite{ldpm1,ldpm2}. Inevitably, there are similarities between the constitutive relations of microplane models and those of LDPM.
 
However, most of the microplane and LDPM simulations have been focusing on the mechanical behavior of isotropic quasi-brittle materials, and there are only a few studies on the formulation for anisotropic quasi-brittle materials \cite{12,14,cu1,clay}. Brocca et al. proposed a microplane formulation for stiff foam based on the assumption that the elastic moduli on the microplanes vary ellipsoidally as function of the microplane orientation \cite{14}. This approach cannot be used to correctly represent the mechanical properties of strongly anisotropic materials due to the fact that an exact correspondence in elasticity between
tensorial macro-stiffness and vectorial micro-stiffness cannot be obtained. A similar limitation exists in the microplane models developed for anisotropic clay \cite{clay, 12}. As shown by Cusatis et al. \cite{cu1}, only the microplane formulation based
on spectral decomposition of the stiffness tensor guarantees that an exact correspondence in elasticity between the microplane formulation and the tensorial formulation can be established. Although the spectral stiffness microplane model is the only known exact and rigorous approach for the anisotropic generalization of the microplane model, such a method becomes less appropriate for the simulation of the nonlinear and softening behaviors of quasi-brittle materials. Cusatis et al. managed to simulate the strain-softening damage and fracture mechanics aspects by using strain-dependent limits to provide bounds for the microplane stresses in each spectral mode \cite{cu1}, but it is not as convenient as directly using microplane normal stress and strain components. This, along with the fact that too many parameters need to be identified in the calibration procedure, renders the method unwieldy in practice, and therefore there remains a scientific challenge to relate
the macroscopic response of anisotropic quasi-brittle materials to the elastic
properties of its underlying microstructure.  

Even for the simulation of the elasticity of isotropic quasi-brittle materials, as it was pointed out by Ba\v zant et al. \cite{6} and Cusatis et al. \cite{ldpm1}, the microplane formulation without volumetric-deviatoric split of the strain cannot cover the entire range of thermodynamically acceptable Poisson's ratios ($-1\leq\nu\leq0.5$): the Poisson's ratio is restricted to the range from $-1$ to $0.25$. Although the full Poisson's ratio range can be obtained by introducing the volumetric-deviatoric decomposition of the normal strain \cite{xx}, this complicates severely the damage formulation. The same issue exists for the simulation of anisotropic quasi-brittle materials, and there still exists no microplane model without spectral decomposition that is capable of giving any thermodynamically admissible Poisson's ratio. In this study, the possibility of formulating a microplane model for transversely isotropic quasi-brittle materials based on the assumption that the elastic moduli on the microplanes vary with the microplane orientation is investigated in details, and the ranges of the Poisson's ratios produced by the model are compared with the full Poisson's ratio range obtained from the thermodynamic restrictions.

\section{Thermodynamic Restrictions on Elastic Constants of Transversely Isotropic Materials}
The elastic stress-strain relation of an anisotropic material can be written in tensorial notation as $\sigma_{ij}=E_{ijkl}\epsilon_{kl}$, where the indices refer to Cartesian coordinates $x_i$ ($i=1, 2, 3$); $\sigma_{ij}$ and $\epsilon_{ij}$ are the second-order stress and strain
tensors, respectively. They are symmetric and their symmetry enables their contraction into six-dimensional vectors $\pmb{\sigma}$ and $\pmb{\epsilon}$. Similarly, the internal and external symmetries of the fourth-order stiffness tensor $E_{ijkl}$ allow its
contraction into a $6 \times 6$ matrix $\bf{E}$. The following rules contract a pair of indices into a single index: $11\to1$, $22\to2$,
$33\to3$, $(23,32)\to4$, $(13,31)\to5$, and $(12,21)\to6$. Therefore, in matrix notation one can write $\pmb{\sigma} = \bf{ E } \pmb{\epsilon}$ where $\pmb{\sigma}=[\sigma_{11}, \sigma_{22}, \sigma_{33}, \sqrt{2}\sigma_{23}, \sqrt{2}\sigma_{13}, \sqrt{2}\sigma_{12}]^T$, $\pmb{\epsilon}=[\epsilon_{11}, \epsilon_{22}, \epsilon_{33}, \sqrt{2}\epsilon_{23}, \sqrt{2}\epsilon_{13}, \sqrt{2}\epsilon_{12}]^T$, and the matrix $\textbf{E}$ is defined accordingly. The foregoing definitions of six-dimensional vectors are known as the
Kelvin notation \cite{kn}. The factor $\sqrt{2}$ assures that both the stiffness tensor and its column matrix have the same
norm, given by the sum of the squares of their elements.

As an anisotropic
medium of the highest symmetry, i.e., hexagonal symmetry, transversely isotropic medium possesses a rotational
symmetry axis and the least number of independent elastic constants (five in total). For the case of transverse isotropy with isotropy in the $1$-$2$ plane, as shown in Fig.~\ref{f1}(a), the elastic compliance matrix, $\bf{C}=\bf{E}^{-1}$, is given by: 
\begin{eqnarray}
\bf{C}=\left[
\begin{array}{cccccc}
1/E & -\nu/E& -\nu'/E'& 0 &0 &0\\
-\nu/E&1/E& -\nu'/E'& 0 &0 &0\\
-\nu'/E' & -\nu'/E'& 1/E'& 0 &0 &0\\
0 & 0& 0& 1/(2G) &0 &0\\
0 & 0& 0& 0 &1/(2G) &0\\
0 & 0& 0& 0 &0 &(1+\nu)/E \end{array} \right] 
\end{eqnarray}
where $E'$, $E$ are Young's moduli in the longitudinal and transverse directions, respectively, $G$ is out-of-plane shear modulus, $\nu'$, $\nu$ are Poisson ratios in the longitudinal and transverse directions, respectively.

It is well known that a necessary and sufficient condition for the work done on an elastic material to be strictly positive is that the matrix $\bf{C}$ be symmetric and positive definite \cite{g}. If the work done on a material is not positive, then useful work could be extracted from the material. This would be a violation of established thermodynamic principles. A necessary and sufficient condition for a symmetric matrix to be positive definite is that all determinants formed from it be positive. In the case of a transversely isotropic material, applying the conditions of positive definiteness to the compliance matrix $\bf{C}$, one can show that the following inequalities must be satisfied by the elastic constants: 
\begin{eqnarray}
\label{2c}
-1<\nu<1\\
\label{2a}
-\sqrt{E'/E}<\nu'<\sqrt{E'/E}\\
\label{2b}
\nu<1-2(\nu')^2({E/E'})
\end{eqnarray}
By making use of these equations, one can find the lower and upper bounds of $\nu'$ for every possible value of $\nu$ with different values of $E/E'$, as plotted in Fig.~\ref{f3}. 

\section{Microplane Model Formulation with Different Types of Constraints}
At the microstructural level, nonlinear and inelastic phenomena often occur on planes of a certain
specific orientation, and thus the constitutive law characterizing the mechanical behavior is best described by
a relationship between stress and strain vectors acting on a generic plane of arbitrary spatial orientation. These microplanes can be imagined as the tangent planes of a unit sphere surrounding every point in
the three-dimensional space \cite{cu1}.

\begin{figure}[b]
  \begin{center}
 \includegraphics[scale=0.35]{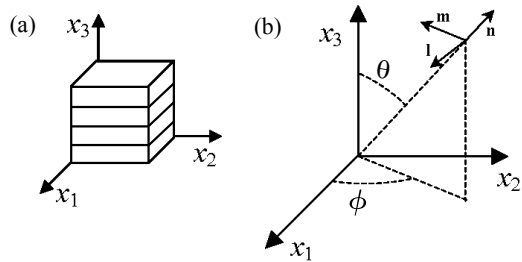}
  \caption{(a) Coordinate system for transversely isotropic materials; and (b) spherical coordinate system.}
  \label{f1}
  \end{center}
\end{figure}

There are two different classes of microplane models: the kinematically constrained and the statically constrained \cite{14}. In the kinematically constrained microplane model, the strain vector on each microplane is the projection of the macroscopic strain tensor. Using Kelvin notation, one can
write $\pmb{\epsilon}_P=\bf{P}\pmb{\epsilon}$ where $\pmb{\epsilon}_P=[\epsilon_{N}, \epsilon_{M}, \epsilon_{L}]^T$ is the microplane strain vector, with $\epsilon_{N}$ being the normal strain component, and $\epsilon_{M}$ and $\epsilon_{L}$ being the shear strain components, respectively. The matrix $\bf{P}$ can be written as:
\begin{eqnarray}
\label{ff3ff}
\bf{P}=\left[
\begin{array}{cccccc}
N_{11} & N_{22}& N_{33}& \sqrt{2}N_{23} &\sqrt{2}N_{13} &\sqrt{2}N_{12}\\
M_{11}&M_{22}&M_{33}& \sqrt{2}M_{23} &\sqrt{2}M_{13} &\sqrt{2}M_{12}\\
L_{11} & L_{22}& L_{33}& \sqrt{2}L_{23} &\sqrt{2}L_{13} &\sqrt{2}L_{12} \end{array} \right] 
\end{eqnarray}
which collects the components of the tensors $N_{ij} = n_in_j$, $M_{ij} = (m_in_j + m_jn_i)/2$ and $L_{ij} = (l_in_j + l_jn_i)/2$,
where $n_i$, $m_i$, and $l_i$ are local Cartesian coordinate vectors on the generic microplane, with $n_i$ being normal. If the
microplane orientation is defined by spherical angles $\theta$ and $\phi$, as shown in Fig.~\ref{f1}(b), then $n_1 = \sin \theta \cos \phi$, $n_2 = \sin \theta \sin \phi$, and $n_3 = \cos \theta$, and one can choose $m_1 = \cos \theta \cos \phi$, $m_2 = \cos \theta \sin \phi$, and $m_3 = -\sin \theta$, which gives $l_1 = -\sin \phi$,
$l_2 = \cos \phi$, and $l_3 = 0$. Once the strain components on each microplane are obtained, the stress components are updated through microplane constitutive laws, which can be expressed in an algebraic or differential form. If the kinematic constraint is imposed, in general, the microplane stress components do not coincide with the projections of the macroscopic stress tensor, i.e., $\pmb{\sigma}_P\neq\textbf{P}\pmb{\sigma}$. Thus static equivalence or equilibrium between the microplane stress components and macroscopic stress tensor must be enforced by other means. This is accomplished by applying the principle of virtual work, which leads to:
\begin{equation}
\pmb{\sigma}=\frac{3}{2\pi}\int_\Omega\textbf{P}^T\pmb{\sigma}_Pd \Omega
\label{2zzhc}
\end{equation}
where $\Omega$ is the surface of a unit hemisphere.

It is possible to formulate the microplane model such that a kinematic constraint for the strains coexists with a static constraint for the stresses. When this happens, the model is said to have a double constraint. As proved by Cusatis et al. \cite{cu1}, such a double constraint exists in the elastic regime if and only if microplane elasticity is formulated through the spectral decomposition of the stiffness or compliance matrices.

\section {Spectral Stiffness Microplane Model}
By using the spectral decomposition theorem \cite{cu1,de1,de2,de3}, the stiffness matrix $\bf{E}$ can be decomposed as $\textbf{E}=\sum\limits_I\lambda_I\textbf{E}_I$ where $\lambda_I$ are the eigenvalues of $\bf{E}$, and $\bf{E}_I$ define a set of matrices constructed from the eigenvectors of $\bf{E}$ as $\textbf{E}_I=\sum\limits_n\pmb{\phi}_{In}\pmb{\phi}_{In}^T$ where $\pmb{\phi}_{In}$ is the normalized eigenvector associated with the eigenvalue $\lambda_I$ of multiplicity $n$ so that $\pmb{\phi}_{In}^T\bf{E}\pmb{\phi}_{In}=\lambda_I$. The following conditions hold for the matrix $\bf{E}$: $\sum\limits_I\textbf{E}_I=1$, $\textbf{E}_I\textbf{E}_I=\textbf{E}_I$, and $\textbf{E}_I\textbf{E}_J=0$ ($I\neq J$). $\textbf{E}_I$ decomposes the stress and strain vectors into energetically orthogonal modes, which are called eigenstresses and eigenstrains, as $\pmb{{\epsilon}}_I=\textbf{E}_I\pmb{\epsilon}$ and $\pmb{{\sigma}}_I=\textbf{E}_I\pmb{\sigma}$, respectively, where $\pmb{{\sigma}}=\sum\limits_I\pmb{{\sigma}}_I$, $\pmb{{\epsilon}}=\sum\limits_I\pmb{{\epsilon}}_I$, and $\pmb{{\sigma}}_I=\lambda_I\pmb{\epsilon_I}$. In a similar manner, one can also decompose the stress and strain vectors into  microplane eigenstresses and microplane eigenstrains as $\pmb{{\epsilon}}_{PI}=\textbf{P}_I\pmb{\epsilon}$ and $\pmb{{\sigma}}_{PI}=\textbf{P}_I\pmb{\sigma}$, respectively, where $\textbf{P}_I=\textbf{P}\textbf{E}_I$ \cite{cu1}. Finally, in the elastic regime, the microplane eigenstresses are proportional to the microplane eigenstrains through the associated eigenvalue, that is, $\pmb{{\sigma}}_{PI}=\lambda_I\pmb{{\epsilon}}_{PI}$.

For the case of transverse isotropy, the eigenvalues of the compliance matrix $\textbf{C}$, which are the reciprocal of the eigenvalues $\lambda_I$ of the
stiffness matrix $\textbf{E}$, can be expressed as \cite{de3,cu1}: $\lambda_1^{-1}=\displaystyle{\frac{1+\nu}{E}}$, $\lambda_2^{-1}=\displaystyle{\frac{1-\nu}{2E}+\frac{1}{2E'}-{\Big[\Big(\frac{1-\nu}{2E}-\frac{1}{2E'}\Big)^2+\frac{2\nu'^2}{E'^2}\Big]}^{1/2}}$, $\lambda_3^{-1}=\displaystyle{\frac{1-\nu}{2E}+\frac{1}{2E'}+{\Big[\Big(\frac{1-\nu}{2E}-\frac{1}{2E'}\Big)^2+\frac{2\nu'^2}{E'^2}\Big]}^{1/2}}$, and $\lambda_4^{-1}=\displaystyle{\frac{1}{2G}}$, and the corresponding $\textbf{E}_I$ can be expressed as:
\begin{equation}
\label{ff1}
\textbf{E}_1=\left[
\begin{array}{cccccc}
 1/2  &  -1/2    &0 &        0    &     0 &        0\\
    -1/2 &  1/2  &  0&         0  &       0&        0\\
    0  &  0  & 0 &        0 &        0 &        0\\
         0   &      0  &       0  & 0 &        0   &      0\\
         0    &     0  &       0   &      0 &  0    &     0\\
         0     &    0 &        0    &     0 &        0  & 0
\end{array} \right]
\end{equation}
\begin{equation}
\textbf{E}_2=\left[
\begin{array}{cccccc}
 c^2/2  &  c^2/2    & cs/\sqrt{2} &        0    &     0 &        0\\
    c^2/2&  c^2/2  &  cs/\sqrt{2}&         0  &       0&        0\\
    cs/\sqrt{2}  &  cs/\sqrt{2} & s^2 &        0 &        0 &        0\\
         0   &      0  &       0  & 0 &        0   &      0\\
         0    &     0  &       0   &      0 &  0    &     0\\
         0     &    0 &        0    &     0 &        0  & 0
\end{array} \right]
\end{equation}
\begin{equation}
\textbf{E}_3=\left[
\begin{array}{cccccc}
 s^2/2  &  s^2/2    & -cs/\sqrt{2} &        0    &     0 &        0\\
    s^2/2&  s^2/2  &  -cs/\sqrt{2}&         0  &       0&        0\\
    -cs/\sqrt{2}  &  -cs/\sqrt{2} & c^2 &        0 &        0 &        0\\
         0   &      0  &       0  & 0 &        0   &      0\\
         0    &     0  &       0   &      0 &  0    &     0\\
         0     &    0 &        0    &     0 &        0  & 0
\end{array} \right]
\end{equation}
\begin{equation}
\label{ff2}
\textbf{E}_4=\left[
\begin{array}{cccccc}
0  & 0    &0 &        0    &     0 &        0\\
   0&  0  &  0&         0  &       0&        0\\
    0  &  0  & 0 &        0 &        0 &        0\\
         0   &      0  &       0  & 1 &        0   &      0\\
         0    &     0  &       0   &      0 &  1    &     0\\
         0     &    0 &        0    &     0 &        0  & 0
\end{array} \right]
\end{equation}
where $c=\cos\omega$, $s=\sin\omega$, and $\tan2\omega=[-2\sqrt{2}\nu'/E']/[(1-\nu)/E-1/E']$. 

As a generalization of the volumetric-deviatoric decomposition, the spectral stiffness microplane model is the only exact and rigorous approach for the anisotropic generalization of the microplane model, but it becomes unwieldy for the simulation of the nonlinear and softening behaviors of quasi-brittle materials. This is because various nonliear and softening laws must be formulated for the different spectral modes and for their interaction. Furthermore, the use in the nonlinear regime of the spectral deformation modes that are derived from the elastic stiffness matrix can be questioned from a theoretical point of view. In many cases, it is easier to formulate nonlinear constitutive equations, especially for fracture and damage, with reference to the total microplane stresses and strains. To directly use normal stress and strain components, a microplane formulation based on the assumption that the elastic moduli on the microplanes vary with the microplane orientation is more convenient. This type of elastic formulation is discussed in the next section.

\section{Microplane Model with Orientation Dependent Moduli}

To capture the macroscopic response of anisotropic materials, Brocca et al. \cite{14} proposed a microplane formulation based on the assumption that the elastic moduli on the microplanes vary as functions of the microplane orientation, that is, $E_i=E_i(\phi,\theta)$, where subscript $i = N, M, L$ labels the components of the microplane strain and stress vectors. Furthermore, for transversely isotropi materials, one can assume that the moduli are functions of $\theta$ only. By integrating the microplane elastic energy over the unit hemisphere, one can obtain:
\begin{equation}
\frac{1}{2}\pmb{\sigma}^T\textbf{E}^*\pmb{\sigma}=W=\frac{3}{2\pi}\int_\Omega\frac{1}{2}\pmb{\sigma}_P^T\textbf{E}_P\pmb{\sigma}_Pd \Omega=\frac{1}{2}\pmb{\sigma}^T\Big[\frac{3}{2\pi}\int_ \Omega\textbf{P}^T\textbf{E}_P\textbf{P}d \Omega\Big]\pmb{\sigma} \Rightarrow \textbf{E}^*=\frac{3}{2\pi}\int_\Omega \textbf{P}^T \textbf{E}_P \textbf{P} d\Omega
\label{2zzx}
\end{equation}
where $\textbf{E}_p=\textbf{diag}(E_i)$. 

The objective of thie study is to investigate the form of the function of $E_i(\theta)$ which gives the maximum range of Poisson's ratios The following four cases are studied. The first is Case A, characterized by a linear variation with $\theta$:
\begin{eqnarray}
\label{mimi}
E_N=(a_1-a_2)\frac{2}{\pi}\theta+a_2;\;\;\; E_M=(a_3-a_4)\frac{2}{\pi}\theta+a_4;\;\;\; E_L=(a_3-a_4)\frac{2}{\pi}\theta+a_4
\end{eqnarray}
The second case, Case B, makes use of trigonometric functions:
\begin{eqnarray}
E_N=a_1\sin^2\theta+a_2\cos^2\theta; \;\;\;
E_M=a_3\sin^2\theta+a_4\cos^2\theta;\;\;\;
E_L=a_3\sin^2\theta+a_4\cos^2\theta
\end{eqnarray}
The third case, Case C, uses the inverse of the functions in Case B:
\begin{eqnarray}
\label{mida}
E_N=(a_1\sin^2\theta+a_2\cos^2\theta)^{-1};\;\;\;
E_M=(a_3\sin^2\theta+a_4\cos^2\theta)^{-1};\;\;\;
\label{mi}
E_L=(a_3\sin^2\theta+a_4\cos^2\theta)^{-1}
\end{eqnarray}
where $a_i$ ($i=1, 2, 3, 4$) are positive unknown parameters. In both Case A and Case B, $E_N|_{\theta=0}=a_2$, $E_N|_{\theta=\pi/2}=a_1$, $E_L|_{\theta=0}=E_M|_{\theta=0}=a_4$, and $E_L|_{\theta=\pi/2}=E_M|_{\theta=\pi/2}=a_3$. In Case C, instead, one has $E_N|_{\theta=0}=1/a_2$, $E_N|_{\theta=\pi/2}=1/a_1$, $E_L|_{\theta=0}=E_M|_{\theta=0}=1/a_4$, and $E_L|_{\theta=\pi/2}=E_M|_{\theta=\pi/2}=1/a_3$. In all the cases, the condition $a_i>0$ ($i=1, 2, 3, 4$) ensures that $E_i>0$ ($i = N, M, L$).

Finally, the fourth case, Case D, assumes independent modulus values at each microplane orientation. For the most commonly adopted quadrature formula with 37 microplanes \cite{BandO,stroud}, this approach involves the values of $E_N$ and $E_M$ at eight different $\theta$: $\theta_1=0$, $\theta_2=0.1\pi$, $\theta_3=0.157\pi$, $\theta_4=0.25\pi$, $\theta_5=0.304\pi$, $\theta_6=0.391\pi$, $\theta_7=0.4\pi$, and $\theta_8=0.5\pi$. Hence, in this case, Young's moduli, $E$ and $E'$, and Poisson's ratios, $\nu$ and $\nu'$, for the transversely isotropic material depend on 16 parameters: $E_M(\theta_1)$, $E_M(\theta_2)$, $E_M(\theta_3)$, $E_M(\theta_4)$, $E_M(\theta_5)$, $E_M(\theta_6)$, $E_M(\theta_7)$, $E_M(\theta_8)$, $E_N(\theta_1)$, $E_N(\theta_2)$, $E_N(\theta_3)$, $E_N(\theta_4)$, $E_N(\theta_5)$, $E_N(\theta_6)$, $E_N(\theta_7)$, and $E_N(\theta_8)$.

Now let us examine the range of the Poisson's ratios that the proposed microplane formulation can generate in each case. Substituting $E_i(\theta)$ as indicated in Eqns.~(\ref{mimi})--(\ref{mi}) into Eqn.~(\ref{2zzx}), one can obtain the following results: 
\begin{eqnarray}
\label{meiy}
\bf{E}^*=\left[
\begin{array}{cccccc}
E_{11} & E_{12}& E_{13}& 0 &0 &0\\
E_{12} & E_{11}& E_{13}& 0 &0 &0\\
E_{13} & E_{13}& E_{33}& 0 &0 &0\\
0 & 0& 0& E_{44} &0 &0\\
0 & 0& 0& 0 &E_{44} &0\\
0 & 0& 0& 0&0 &E_{66}\\
\end{array}
\right]
\end{eqnarray}
where for Case A, one can obtain the following:
\begin{eqnarray}\label{ffkoko}
E_{11}=[447a_1+3(-149+60\pi)a_2+253a_3+(120\pi-253)a_4]/(300\pi)\\
E_{12}=[149a_1+(-149+60\pi)a_2-149a_3+(149-60\pi)a_4]/(300\pi)\\
E_{13}=[26a_1+(-26+15\pi)a_2-26a_3+(26-15\pi)a_4]/(75\pi)\\
E_{33}=[48a_1+(-48+45\pi)a_2+52a_3+(-52+30\pi)a_4]/(75\pi)
\end{eqnarray}
For Case B, one has the following results: 
\begin{eqnarray}\label{ffkoko2}
E_{11}=(18a_1+3a_2+10a_3+4a_4)/35\\
E_{12}=(6a_1+a_2-6a_3-a_4)/35\\
E_{13}=(4a_1+3a_2-4a_3-3a_4)/35\\
E_{33}=(6a_1+15a_2+8a_3+6a_4)/35
\end{eqnarray}
For Case C, one has the following results: 
\begin{eqnarray}\label{fnf}
&E_{11}=\displaystyle{\frac{3}{8}\Bigg[\frac{2a_3+a_4}{(a_4-a_3)^2}-\frac{(4a_3-a_4){a_4}{a_3^{-1/2}}\arctan[{(a_4-a_3)}^{1/2}a_3^{-1/2}]}{(a_4-a_3)^{5/2}}}\\\nonumber&+\displaystyle{\frac{2a_1-5a_2}{(a_2-a_1)^2}+\frac{3{a_2^2}{a_1^{-1/2}}\arctan[{(a_2-a_1)}^{1/2}a_1^{-1/2}]}{(a_2-a_1)^{5/2}}\Bigg]}\\\label{ff2}
&E_{12}=\displaystyle{\frac{3}{8}\Bigg[\frac{5a_4-2a_3}{3(a_4-a_3)^2}-\frac{{a_4}^2{a_3^{-1/2}}\arctan[{(a_4-a_3)}^{1/2}a_3^{-1/2}]}{(a_4-a_3)^{5/2}}}\\\nonumber&+\displaystyle{\frac{2a_1-5a_2}{3(a_2-a_1)^2}+\frac{{a_2^2}{a_1^{-1/2}}\arctan[{(a_2-a_1)}^{1/2}a_1^{-1/2}]}{(a_2-a_1)^{5/2}}\Bigg]}\\\label{ff3}
&E_{13}=\displaystyle{\frac{1}{2}\Bigg[-\frac{2a_4+a_3}{(a_4-a_3)^2}+\frac{3{a_4}{a_3^{1/2}}\arctan[{(a_4-a_3)}^{1/2}a_3^{-1/2}]}{(a_4-a_3)^{5/2}}}\\\nonumber&+\displaystyle{\frac{2a_2+a_1}{(a_2-a_1)^2}-\frac{3{a_2}{a_1^{1/2}}\arctan[{(a_2-a_1)}^{1/2}a_1^{-1/2}]}{(a_2-a_1)^{5/2}}\Bigg]}\\\label{ff4}
&E_{33}=\displaystyle{\frac{2a_4+a_3}{(a_4-a_3)^2}-\frac{3{a_4}{a_3^{1/2}}\arctan[{(a_4-a_3)}^{1/2}a_3^{-1/2}]}{(a_4-a_3)^{5/2}}}\\\nonumber&+\displaystyle{\frac{a_2-4a_1}{(a_2-a_1)^2}+\frac{3{a_1^{3/2}}\arctan[{(a_2-a_1)}^{1/2}a_1^{-1/2}]}{(a_2-a_1)^{5/2}}}
\end{eqnarray} 

Note that Eqns.~(\ref{fnf})-(\ref{ff4}) are valid only when $a_2>a_1$ and $a_4>a_3$. When $a_2<a_1$, $\arctan[(a_2-a_1)^{1/2} a_1^{-1/2}](a_2-a_1)^{-5/2}$ needs to be replaced by $\displaystyle{{\text{arctanh}[{(a_1-a_2)}^{1/2}a_1^{-1/2}]}{(a_1-a_2)^{-5/2}}}$; and similarly, when $a_4<a_3$, $\displaystyle{{\arctan[{(a_4-a_3)}^{1/2}a_3^{-1/2}]}{(a_4-a_3)^{-5/2}}}$ needs to be replaced by $\{\text{arctanh}[{(a_3-a_4)}^{1/2}a_3^{-1/2}](a_3-a_4)^{-5/2}$ in Eqns.~(\ref{fnf})-(\ref{ff4}).

Young's moduli and Poisson's ratios for transversely isotropic materials can be written as:
\begin{eqnarray}
\label{ff00}
E=(E^2_{11}E_{33}+2E_{13}^2E_{12}-2E_{11}E^2_{13}-E_{33}E^2_{12})/(E_{11}E_{33}-E_{13}^2)\\
\label{ff0}
E'=(E^2_{11}E_{33}+2E_{13}^2E_{12}-2E_{11}E^2_{13}-E_{33}E^2_{12})/(E_{11}^2-E_{12}^2)\\
\label{ff}
\nu=(E_{12}E_{33}-E_{13}^2)/(E_{11}E_{33}-E_{13}^2)\\
\label{ffnn}
\nu'=E_{13}/(E_{11}+E_{12})
\end{eqnarray}
Defining the following dimensionless variables: $t=E'/E$, $A=E_{33}/E_{11}$, $B=E_{13}/E_{11}$, and $C=E_{12}/E_{11}$, and one has:
\begin{eqnarray}
\label{ff00xx}
t=E'/E=(A-B^2)/(1-C^2)\\
\label{ffxx}
\nu=(CA-B^2)/(A-B^2)\\
\label{ffnnxx}
\nu'=B/(1+C)
\end{eqnarray}

\begin{figure}[b]
  \begin{center}
 \includegraphics[height=13cm,width=4.78cm]{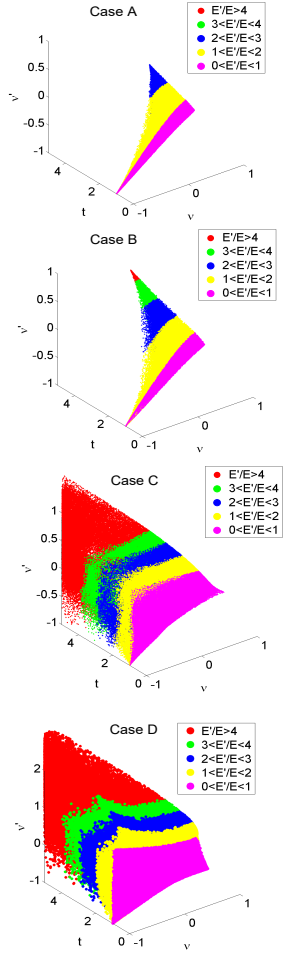}
  \caption{The values of $t(\alpha, \beta, \gamma)$, $\nu(\alpha, \beta, \gamma)$, and $\nu'(\alpha, \beta, \gamma)$ for any $\alpha>0$, $\beta>0$, and $\gamma>0$ for each case with different values of $t$ indicated by different colors. The figure appears in color in the electronic version of this article.}
  \label{f2}
  \end{center}
\end{figure}

Furthermore, by setting $\alpha=a_2/a_1$, $\beta=a_3/a_1$, and $\gamma=a_4/a_1$, one can plot the values of $t(\alpha, \beta, \gamma)$, $\nu(\alpha, \beta, \gamma)$, and $\nu'(\alpha, \beta, \gamma)$ for any $\alpha>0$, $\beta>0$, and $\gamma>0$. Calculated from $10^8$ randomly generated positive real numbers used as $\alpha$, $\beta$, or $\gamma$, the results for each case are shown in Fig.~\ref{f2} with different values of $t$ indicated by different colors. (The figures appear in color in the electronic version of this article.) It can be seen that the ranges of $\nu$, $\nu'$, and $t$ generated by Case B are only slightly larger than those obtained from Case A, but significantly smaller than those obtained from Case C and Case D.

\begin{figure}[b]
  \begin{center}
 \includegraphics[scale=0.22]{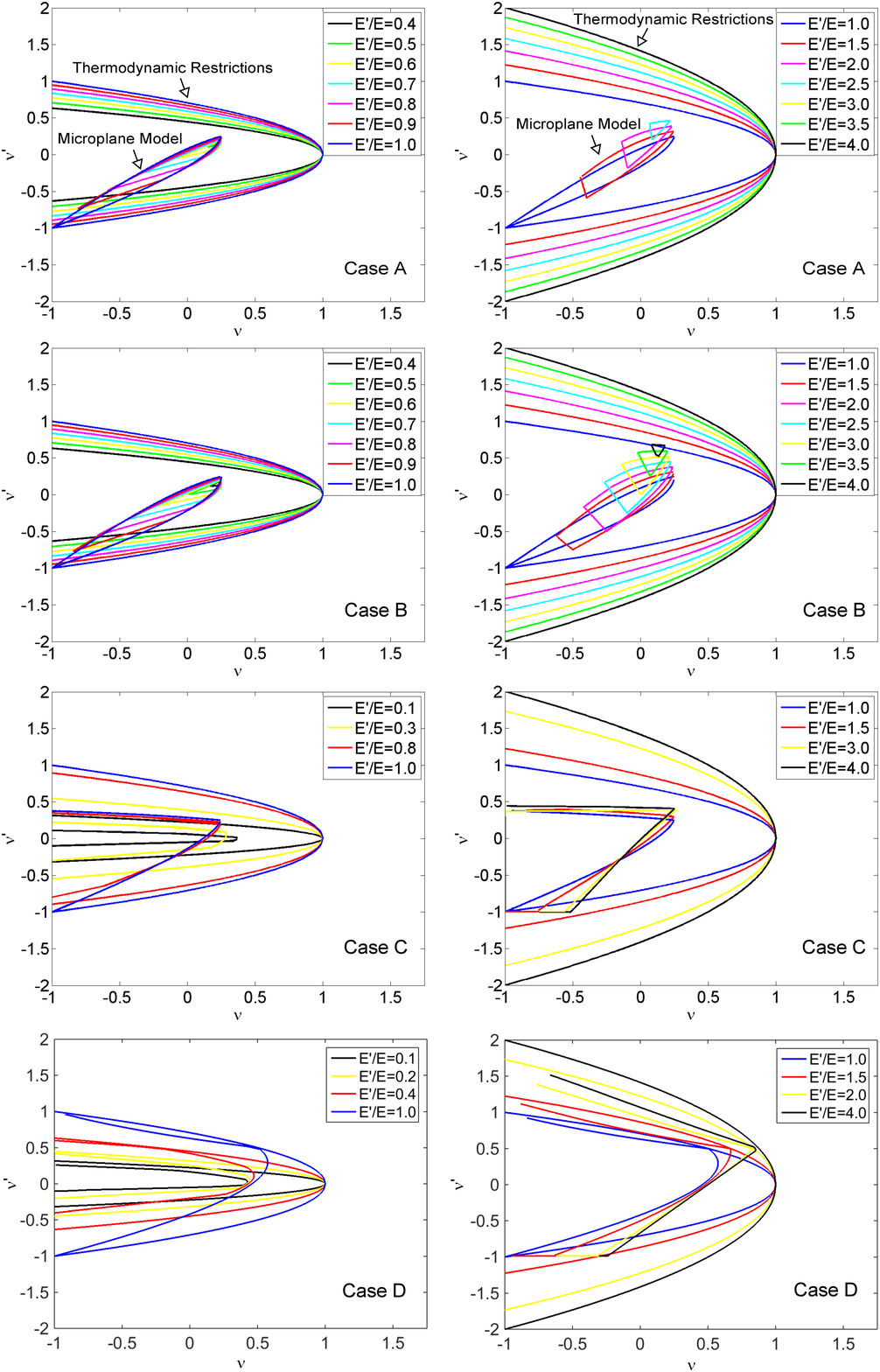}
  \caption{The contour plot of $t$ for each case. The results are compared with the thermodynamic restrictions on elastic constants of transversely isotropic materials. The figure appears in color in the electronic version of this article.}
  \label{f3}
  \end{center}
\end{figure}

To further confirm that Case C is the best scenario, one can obtain the contour plot of $t$ for each case. Substituting Eqns.~(\ref{ff00xx}) and (\ref{ffnnxx}) into Eqn.~(\ref{ffxx}), the function of $\nu=\nu(\nu',t,C)$ can be obtained as follows:
\begin{equation}
\nu=C-\nu'^2(1+C)/t
\end{equation}
To obtain the upper and lower bounds of $\nu$ for every possible value of $\nu'$ with different values of $t$, one needs to maximize and minimize $v(\alpha, \beta, \gamma)$, for any $\alpha>0$, $\beta>0$, and $\gamma>0$, subject to the constraints that $\nu'(\alpha, \beta, \gamma) = \nu'_0$ and $t(\alpha, \beta, \gamma)=t_0$. The results for each case are shown in Fig.~\ref{f3} with different values of $t$ indicated by different colors, and they are compared with the thermodynamic restrictions on elastic constants of transversely isotropic materials obtained from Eqns.~(\ref{2c})-(\ref{2b}). Since Fig.~\ref{f3} confirms that Case C and Case D generate the largest ranges of $\nu$, $\nu'$, and $t$, they will be adopted for the numerical modeling of transversely isotropic elasticity of quasi-brittle materials in this study. Two examples are given as shown below.

\subsection {Elastic Microplane Model Formulation for Shale} 
Adequate knowledge and prediction of mechanical properties of shale are pivotal to the success in many fields of petroleum engineering, ranging from seismic exploration, to well drilling and production, and to the design of hydraulic fractures. Shale is best described as transversely isotropic quasi-brittle material with the symmetric axes being perpendicular to bedding. In laboratory measurements of shale, high magnitude of anisotropy was reported for both static \cite{shale1} and dynamic \cite{shale2} conditions, which cannot be neglected in shale modeling.  Neglecting shale anisotropy may lead to incorrect estimates of rock and fluid properties, fracture aperture, fracture containment, and stress or stress changes resulting from production. To our knowledge, a microplane model to completely characterize the transversely isotropic elastic behavior of shale has not yet been developed. 

\begin{figure}[b]
  \begin{center}
 \includegraphics[height=7.53cm,width=19cm]{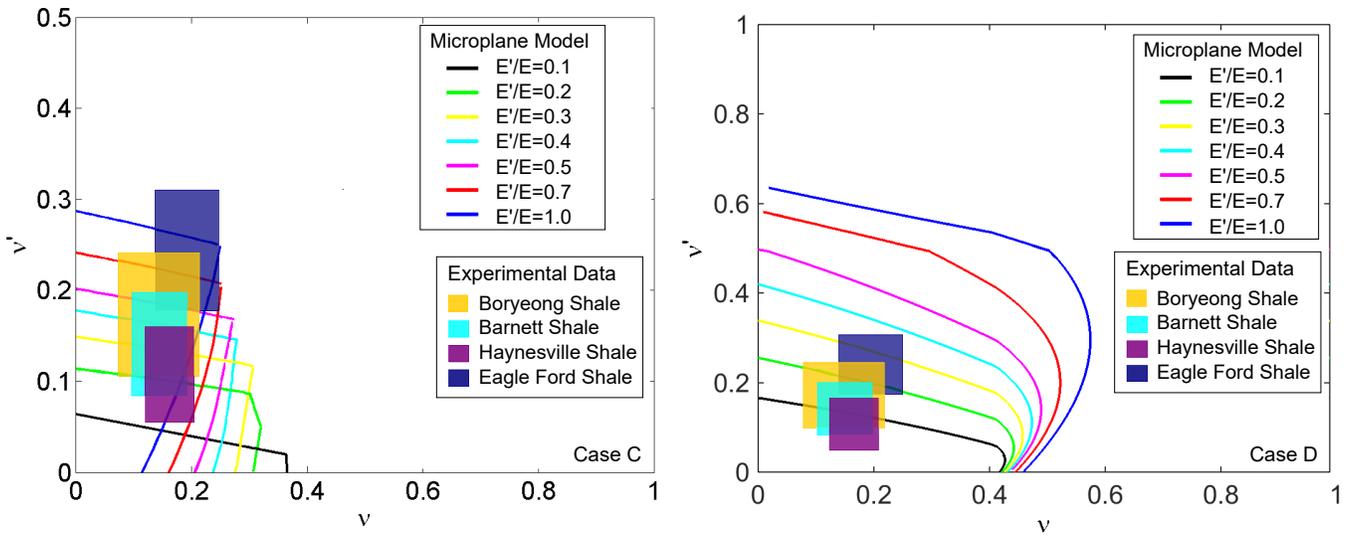}
  \caption{The ranges of $\nu$ and $\nu'$ obtained from microplane model based on Case C and Case D when $0<E'/E\le 1$, respectively. The ranges of $\nu$ and $\nu'$ for various types of shale studied by existing literature \cite{1j,2j,3j,5j,55j,6j} are also plotted. The figure appears in color in the electronic version of this article.} 
  \label{f4}
  \end{center}
\end{figure}

Due to the presence of bedding-parallel weakness planes, shales are in general stiffer along the bedding planes than perpendicular to bedding, i.e, $E'/E<1$. Fig.~\ref{f4} plots the ranges of $\nu$ and $\nu'$ obtained from microplane model based on Case C for seven different values of $E'/E$ when $0<E'/E\le 1$. The ranges of $\nu$ and $\nu'$ for various types of shale provided by existing literature \cite{1j,2j,3j,5j,55j,6j} are also plotted on Fig.~\ref{f4}. It can be seen that the ranges of $\nu$ and $\nu'$ for most types of shale fall within the microplane simulation region. The ranges of $\nu$ and $\nu'$ based on Case D are also plotted in Fig.~\ref{f4} for comparison. Again, it shows that the possible range of Poisson's ratios obtained from Case D are much larger than those obtained from Case C.
 
One can take Boryeong shale as an example, which has been extensively investigated in the literature. The experimental data on the five elastic constants of Boryeong shale are provided by Cho et. al. \cite{3j}, as shown in Table 1. Based on the experimental data, one has $E=37.3$ GPa, $E'=18.4$ GPa, $\nu=0.15$, $\nu'=0.16$, and $G=12.0$ GPa, and the elastic stiffness matrix, $\bf{E}$, reads: 
\begin{eqnarray}
\textbf{E}=\left[
\begin{array}{cccccc}
 41.2104  &  8.7756    &7.9978 &        0    &     0 &        0\\
    8.7756 &  41.2104  &  7.9978&         0  &       0&        0\\
    7.9978  &  7.9978  & 20.9593 &        0 &        0 &        0\\
         0   &      0  &       0  & 24.0000 &        0   &      0\\
         0    &     0  &       0   &      0 &  24.0000    &     0\\
         0     &    0 &        0    &     0 &        0  & 32.4348
\end{array} \right] \text{GPa}
\end{eqnarray}

\begin{figure}[b]
  \begin{center}
 \includegraphics[scale=.22]{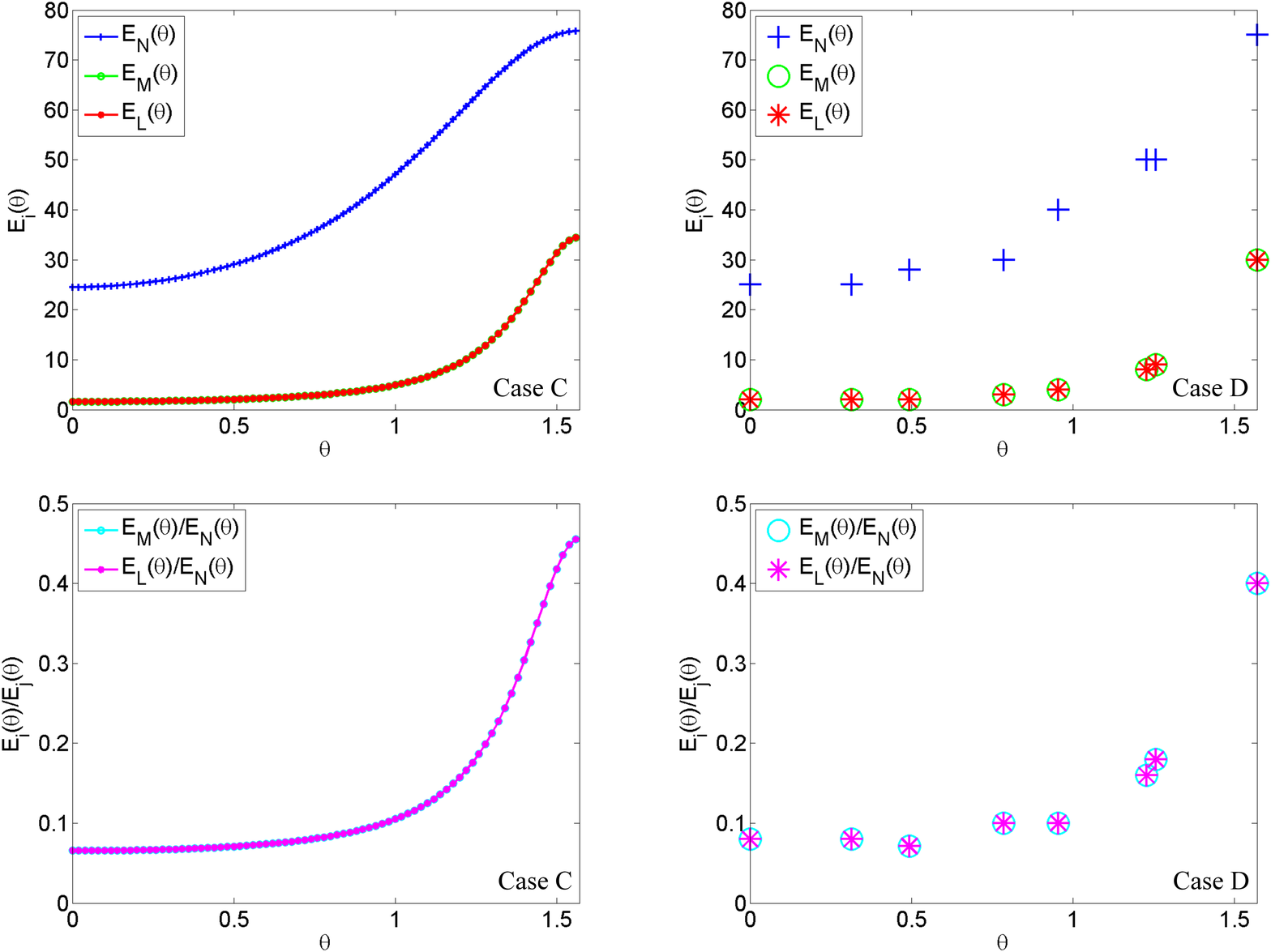}
  \caption{The results for the values and the ratios of $E_i$ ($i = N, M, L$) as a function of $\theta$ obtained from Case C and Case D for Example A, respectively. } 
  \label{f5}
  \end{center}
\end{figure}

\begin{figure}[b]
  \begin{center}
 \includegraphics[scale=.29]{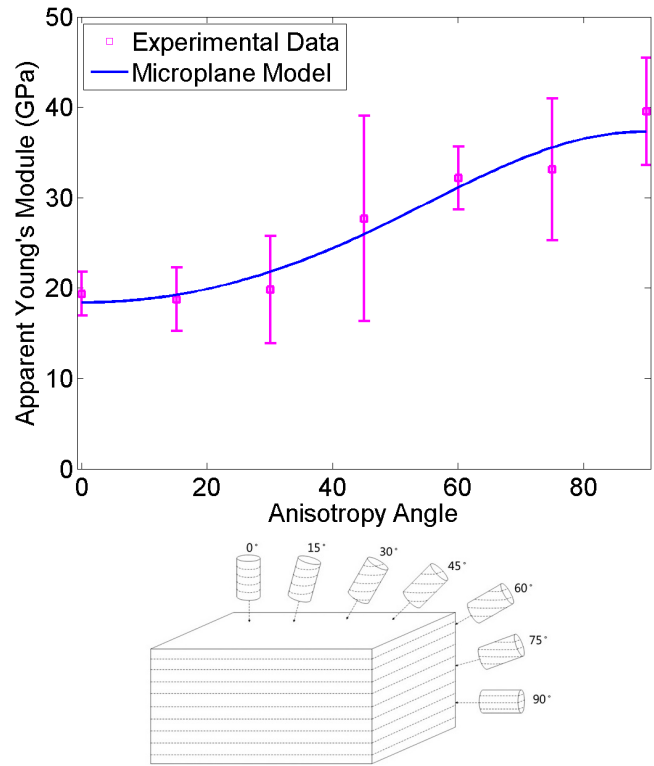}
  \caption{The variation of apparent Young's modulus with anisotropy angle in comparison with experimental data provided by Cho et al. \cite{3j}. } 
  \label{f6}
  \end{center}
\end{figure}

By adopting the formulation provided by Case C, and determining the unknown parameters $a_i$ ($i=1, 2, 3, 4$) by minimizing the Frobenius norm $\sqrt{\sum\limits_{i,j}|E^*_{ij}-E_{ij}|^2}$, where $\bf{E^*}$ is defined in Eqn.~(\ref{2zzx}), one obtains $a_1=0.0132 \text{ GPa}^{-1}$, $a_2=0.0408 \text{ GPa}^{-1}$, $a_3=0.0289 \text{ GPa}^{-1}$, and $a_4=0.6227 \text{ GPa}^{-1}$, which gives the following results:            
\begin{eqnarray}
\textbf{E}^*=\left[
\begin{array}{cccccc}
   41.4836 &   8.8028    &7.5865 &        0      &   0 &        0\\
    8.8028  & 41.4836   & 7.5865  &       0     &    0  &       0\\
    7.5865   & 7.5865  & 20.9401   &      0    &     0   &      0\\
         0    &     0 &        0   &24.1149   &      0    &     0\\
         0     &    0&        0     &    0   &24.1149      &   0\\
         0      &   0&         0     &    0  &       0   &32.0101
\end{array} \right] \text{GPa}
\end{eqnarray}
It can be seen that a good match between $\bf{E}^*$ and $\bf{E}$ has been obtained. Based on Eqn.~(\ref{mida}), one can plot the curves for the values and the ratios of $E_i$ ($i = N, M, L$) as a function of $\theta$, as shown in Fig.~\ref{f5}. Fig.~\ref{f6} shows the variation of apparent Young's modulus with anisotropy angle in comparison with experimental data provided by Cho et al. \cite{3j}. 

Alternatively, Case D can also be applied. The unknown parameters $E_M(\theta_1)$, $E_M(\theta_2)$, $E_M(\theta_3)$, $E_M(\theta_4)$, $E_M(\theta_5)$, $E_M(\theta_6)$, $E_M(\theta_7)$, $E_M(\theta_8)$, $E_N(\theta_1)$, $E_N(\theta_2)$, $E_N(\theta_3)$, $E_N(\theta_4)$, $E_N(\theta_5)$, $E_N(\theta_6)$, $E_N(\theta_7)$, and $E_N(\theta_8)$ can be determined by minimizing the Frobenius norm $\sqrt{\sum\limits_{i,j}|E^*_{ij}-E_{ij}|^2}$. One obtains the following results: 
\begin{eqnarray}
\textbf{E}^*=\left[
\begin{array}{cccccc}
  41.2104  &  8.7756    &7.9978 &        0    &     0 &        0\\
    8.7756 &  41.2104  &  7.9978&         0  &       0&        0\\
    7.9978  &  7.9978  & 20.9593 &        0 &        0 &        0\\
         0   &      0  &       0  & 24.0000 &        0   &      0\\
         0    &     0  &       0   &      0 &  24.0000    &     0\\
         0     &    0 &        0    &     0 &        0  & 32.4348
\end{array} \right] \text{GPa}
\end{eqnarray}
In this case, an exact match is obtained. The results for the values and the ratios of $E_i$ ($i = N, M, L$) as a function of $\theta$ are shown in Fig.~\ref{f5}.           

\subsection {Elastic Microplane Model Formulation for Rigid Polymeric Foams} 
Foamed plastics, such as polyurethane, polyvinyl chloride (PVC), polystyrene, polypropylene, epoxy, phenol-formaldehyde, cellulose acetate, and silicone, are widely used as core materials for sandwich structures in automotive and aerospace industries due to their light weight and high specific stiffness. They are good heat insulators by virtue of the low conductivity of the gas contained in the cells; they have a higher ratio of flexural modulus to density than before foaming; and they achieve a greater load-bearing capacity per unit weight, as well as greater energy storage and energy dissipation capacities \cite{14,foam1,foam3}. 

However, most of the polymeric foams usually show an anisotropic behavior, which complicates the numerical modeling of such materials. For simplicity, the elastic response of polymeric foams is usually regarded as transversely isotropic: during the foaming process, viscous forces cause the cells to be elongated in the rise direction, and therefore the material response in this direction is stiffer, i.e, $E'/E>1$. The ratio of the modulus in the rise direction to that in the
perpendicular-to-rise direction is indicative of the extent of elongation of the cells.

Let's take rigid PVC foams as an example. The experimental data on the five elastic constants of DIAB Divinycell H60 are provided by the DIAB group \cite{foam5} and Tita et. al. \cite{foam4}, as shown in Table 2. Based on the experimental data, one has $E=16.0$ GPa, $E'=32.0$ GPa, $\nu=0.29$, $\nu'=0.28$, and $G=15.0$ GPa, and the elastic stiffness matrix, $\textbf{E}$, reads: 
\begin{eqnarray}
\textbf{E}=\left[
\begin{array}{cccccc}
 18.8678  &  6.4647    &7.0931   &      0    &     0   &      0\\
    6.4647  & 18.8678  &  7.0931 &        0  &       0 &        0\\
    7.0931   & 7.0931  & 35.9721  &       0  &       0 &        0\\
         0    &     0  &       0  & 30.0000  &       0 &        0\\
         0     &    0  &       0   &      0  & 30.0000 &       0\\
         0      &   0 &        0   &      0  &       0 &  12.4031
\end{array} \right] \text{GPa}
\end{eqnarray}

\begin{figure}[b]
  \begin{center}
 \includegraphics[scale=.22]{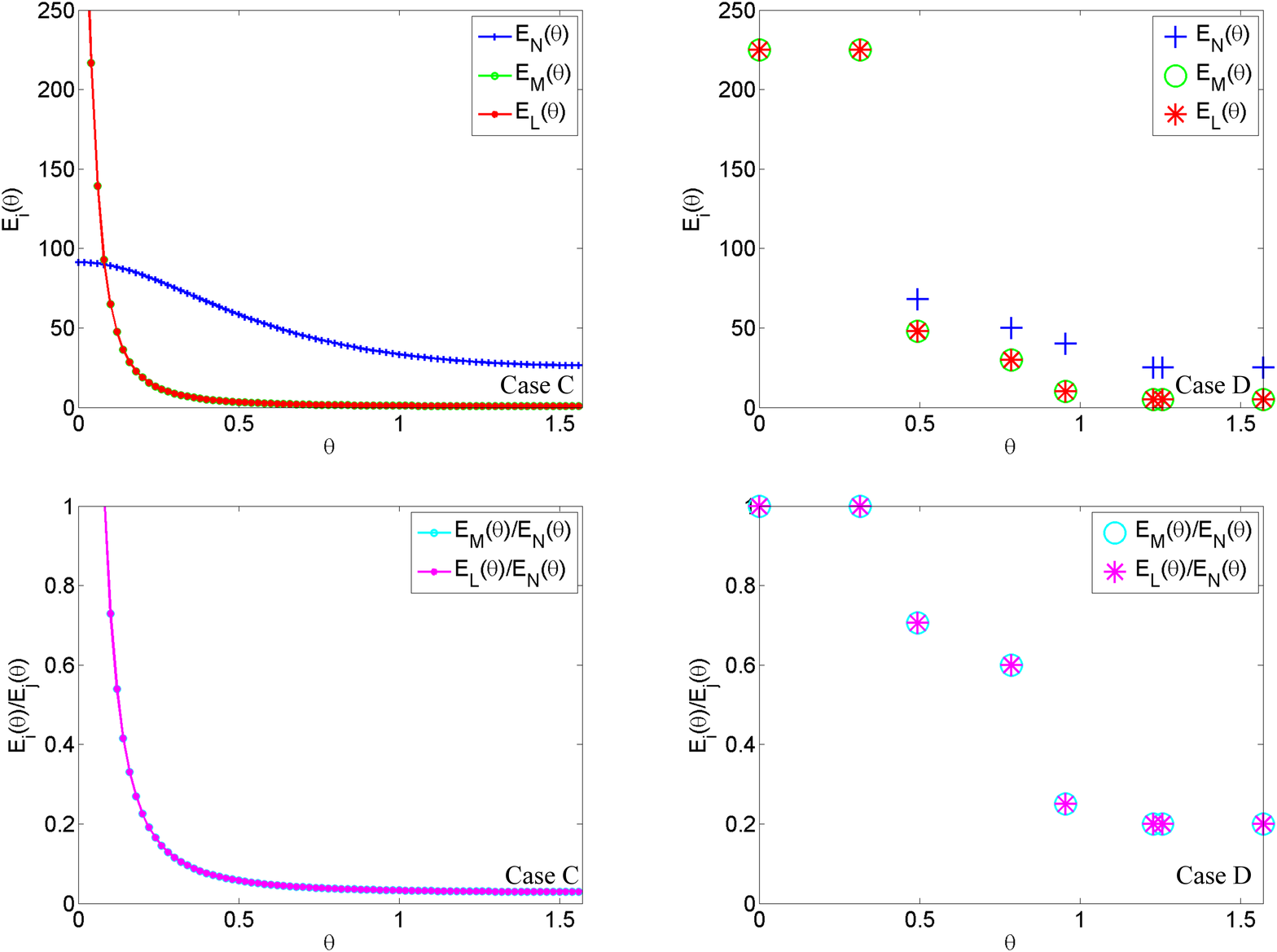}
  \caption{The results for the values and the ratios of $E_i$ ($i = N, M, L$) as a function of $\theta$ obtained from Case C and Case D for Example B, respectively. } 
  \label{f55}
  \end{center}
\end{figure}

By adopting the formulation provided by Case C, and determining the unknown parameters $a_i$ ($i=1, 2, 3, 4$) by minimizing $\sqrt{\sum\limits_{i,j}|E^*_{ij}-E_{ij}|^2}$, one has $a_1=0.0378 \text{ GPa}^{-1}$, $a_2=0.0109 \text{ GPa}^{-1}$, $a_3=1.2882 \text{ GPa}^{-1}$, and $a_4=0.0026 \text{ GPa}^{-1}$, which gives the following results:                           
\begin{eqnarray}
\textbf{E}^*=\left[
\begin{array}{cccccc}
   18.6261  &  5.8339  &  7.7705 &        0   &      0   &      0\\
    5.8339   &18.6261 &   7.7705  &       0  &       0   &      0\\
    7.7705    &7.7705 &  35.9102   &      0  &       0    &     0\\
         0     &    0&         0   &30.0000  &       0     &    0\\
         0      &   0&         0    &     0  & 30.0000      &   0\\
         0       &  0&         0     &    0  &       0 &  12.8046
\end{array} \right] \text{GPa}
\end{eqnarray}
It can be seen that the match between $\bf{E}$ and $\bf{E}^*$ obtained from Case C is not very accurate but still satisfactory. Fig.~\ref{f55} plots the curves for the values and the ratios of $E_i$ ($i = N, M, L$) as a function of $\theta$. 

For Case D, the optimized microplane parameters give:
\begin{eqnarray}
\textbf{E}^*=\left[
\begin{array}{cccccc}
  18.8678  &  6.4647    &7.0931   &      0    &     0   &      0\\
    6.4647  & 18.8678  &  7.0931 &        0  &       0 &        0\\
    7.0931   & 7.0931  & 35.9721  &       0  &       0 &        0\\
         0    &     0  &       0  & 30.0000  &       0 &        0\\
         0     &    0  &       0   &      0  & 30.0000 &       0\\
         0      &   0 &        0   &      0  &       0 &  12.4031
\end{array} \right] \text{GPa}
\end{eqnarray}
which basically coincides with $\bf{E}$. The results for the values and the ratios of $E_i$ ($i = N, M, L$) as a function of $\theta$ are shown in Fig.~\ref{f55}.
 
\section {Comparison between Microplane Model with Orientation Dependent Moduli and Spectral Stiffness Microplane Model}
Note that Eqns.~(\ref{mimi})-(\ref{mi}) are just assumptions on the form of $E_i$ ($i = N, M, L$), and the actual form of $E_i$ can be obtained only when the microplane model is under double constraint, which is derived through the spectral stiffness microplane model. It is worth then studying the accuracy with which the non-spectral formulation approximates the actual microplane stress distribution.

\begin{figure}[b]
  \begin{center}
 \includegraphics[scale=.26]{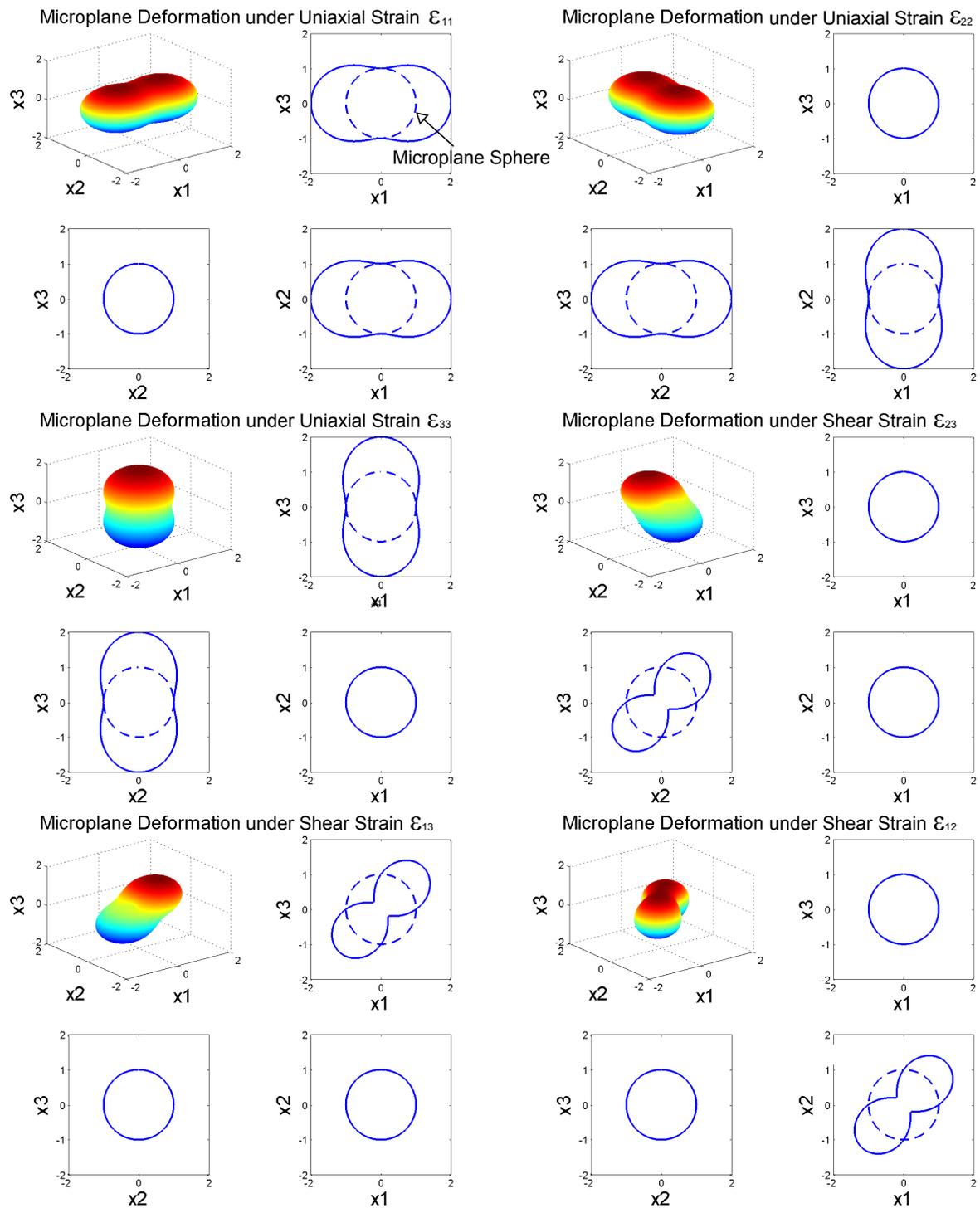}
  \caption{The distribution of the normal strain component, $\epsilon_N$, on a generic microplane sphere caused by different types of macroscopic strains for the Boryeong shale with $E=37.3$ GPa, $E'=18.4$ GPa, $\nu=0.15$, $\nu'=0.16$, and $G=12.0$ GPa.} 
  \label{f7}
  \end{center}
\end{figure}

\begin{figure}[b]
  \begin{center}
 \includegraphics[scale=.26]{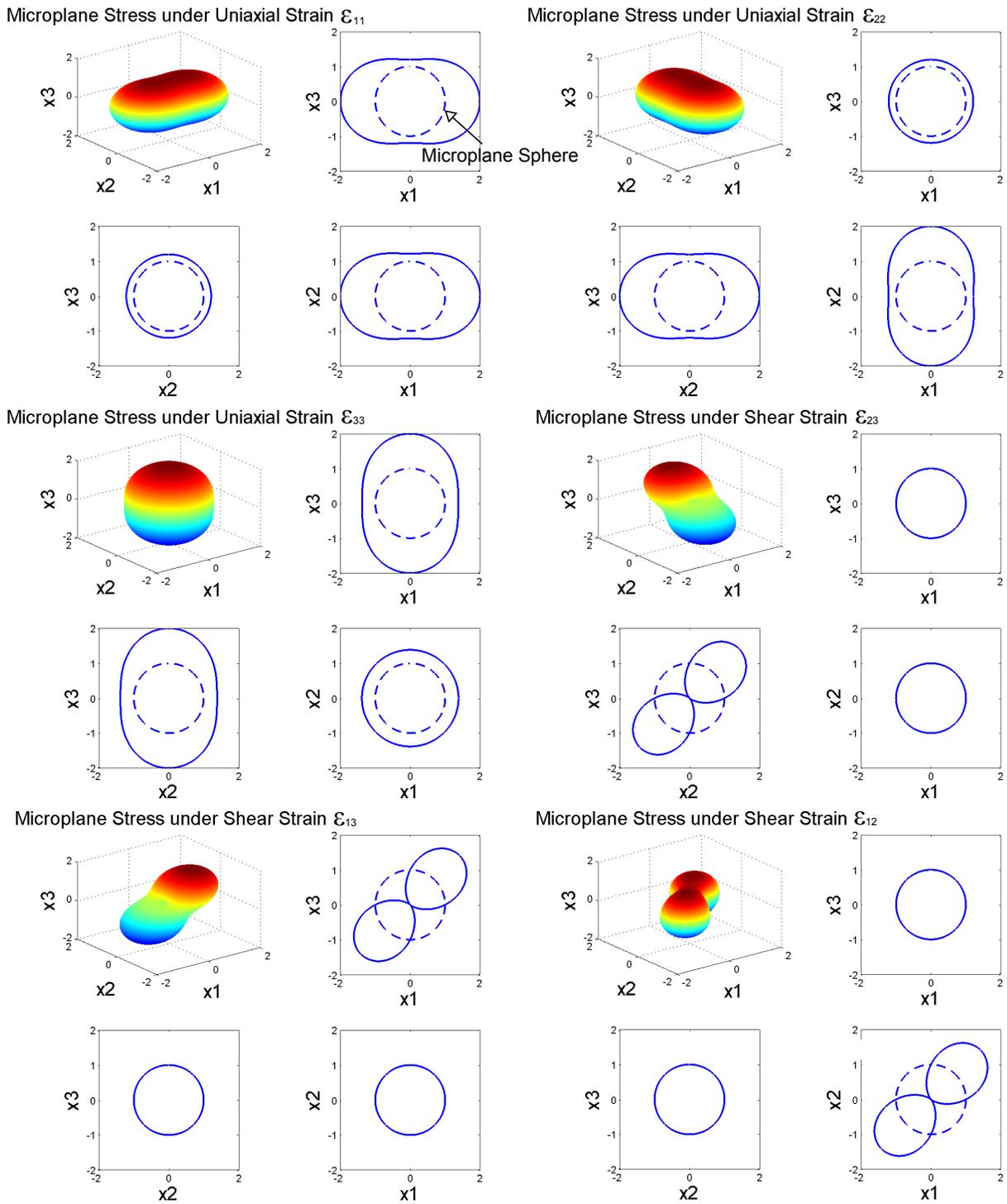}
  \caption{The distribution of the normalized normal stress component, $\sigma_N$, on a generic microplane sphere caused by different types of macroscopic strains for the Boryeong shale with $E=37.3$ GPa, $E'=18.4$ GPa, $\nu=0.15$, $\nu'=0.16$, and $G=12.0$ GPa.} 
  \label{f8}
  \end{center}
\end{figure}

\begin{figure}[b]
  \begin{center}
 \includegraphics[scale=.26]{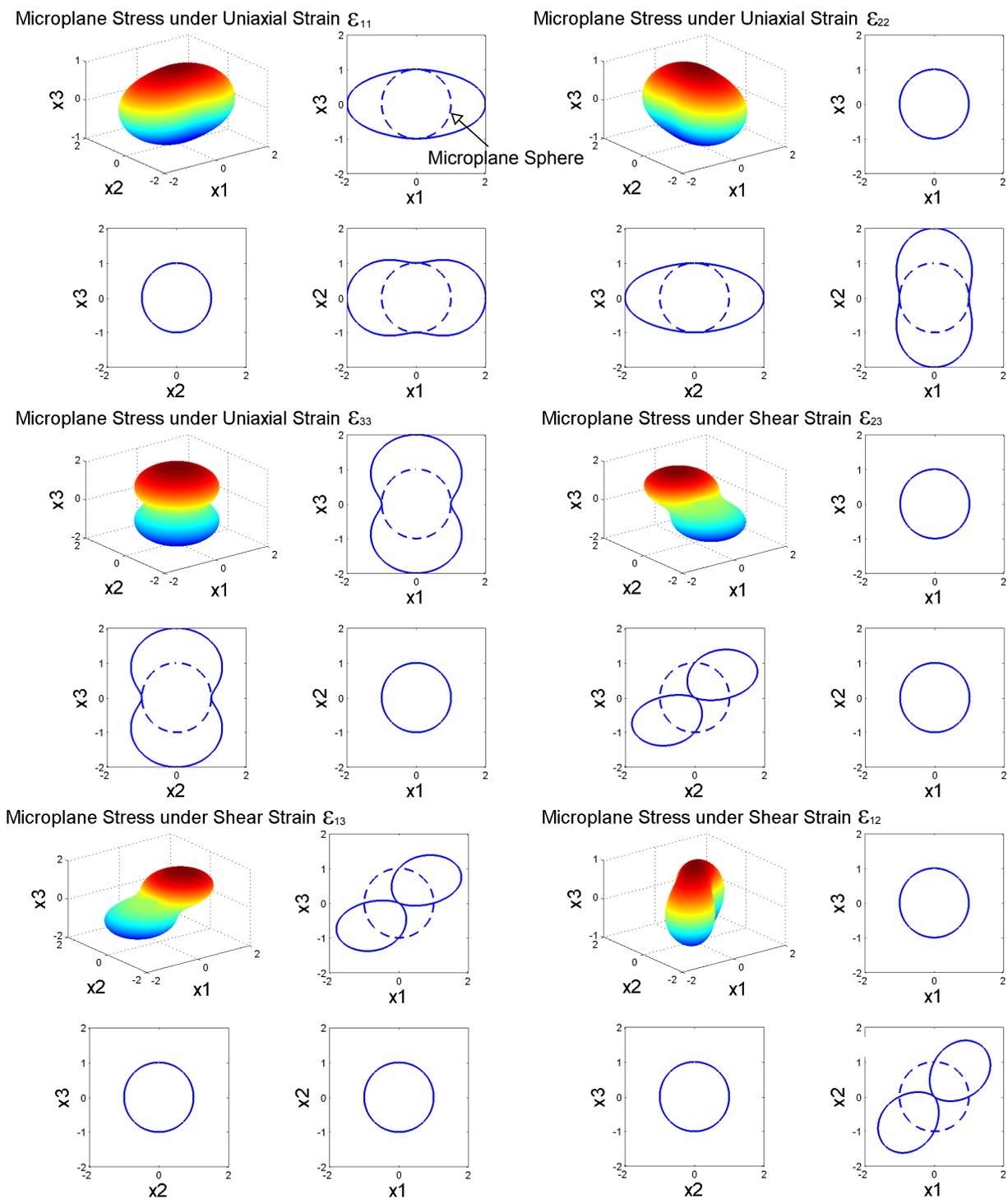}
  \caption{The distributions of the normalized normal stress component $\sigma_N$ based on the assumption that $\sigma_N=E_N\epsilon_N$, where $E_N=1/(a_1\sin^2\theta+a_2\cos^2\theta)$ as given in Case C, caused by different types of macroscopic strains for the Boryeong shale with $E=37.3$ GPa, $E'=18.4$ GPa, $\nu=0.15$, $\nu'=0.16$, and $G=12.0$ GPa.} 
  \label{f9}
  \end{center}
\end{figure}

By taking the Boryeong shale again as reference, the distributions of the normal strain component, $\epsilon_N$, on a generic microplane sphere caused by different types of macroscopic strains are shown in Fig.~\ref{f7}. It has six sub-figures, corresponding to the the distribution of $\epsilon_N$ on the microplane sphere under uniaxial strain $\epsilon_{11}$, $\epsilon_{22}$, and $\epsilon_{33}$, and shear strain $\epsilon_{23}$, $\epsilon_{13}$, and $\epsilon_{12}$, respectively. Each sub-figure includes one three-dimensional plot and three contours plots on the $x_1$-$x_3$ plane, the $x_2$-$x_3$ plane and the $x_1$-$x_2$ plane, respectively. In a similar manner, the distributions of the normalized normal stress component, $\sigma_N$, are shown in Fig.~\ref{f8}. For the purpose of comparison, Fig.~\ref{f9} plots the distributions of the normalized normal stress component $\sigma_N$ based on the assumption that $\sigma_N=E_N\epsilon_N$, where $E_N=1/(a_1\sin^2\theta+a_2\cos^2\theta)$ as given in Case C. It can be seen that, while not an exact match, the stress distribution obtained from the formulation in Case C matches closely with the actual stress distribution. The  deviation of  $\sigma_N$ based on orientation variation microplane model from the one based on spectral stiffness microplane model is typically in the range of 11\% to 28\%: the deviation under uniaxial strain $\epsilon_{11}$ or $\epsilon_{22}$ is less than 20\%; the deviation under uniaxial strain $\epsilon_{33}$ is less than 28\%; and the deviation under shear strain $\epsilon_{23}$, $\epsilon_{13}$, or $\epsilon_{12}$ is less than 11\%.

\begin{figure}[b]
  \begin{center}
 \includegraphics[scale=.26]{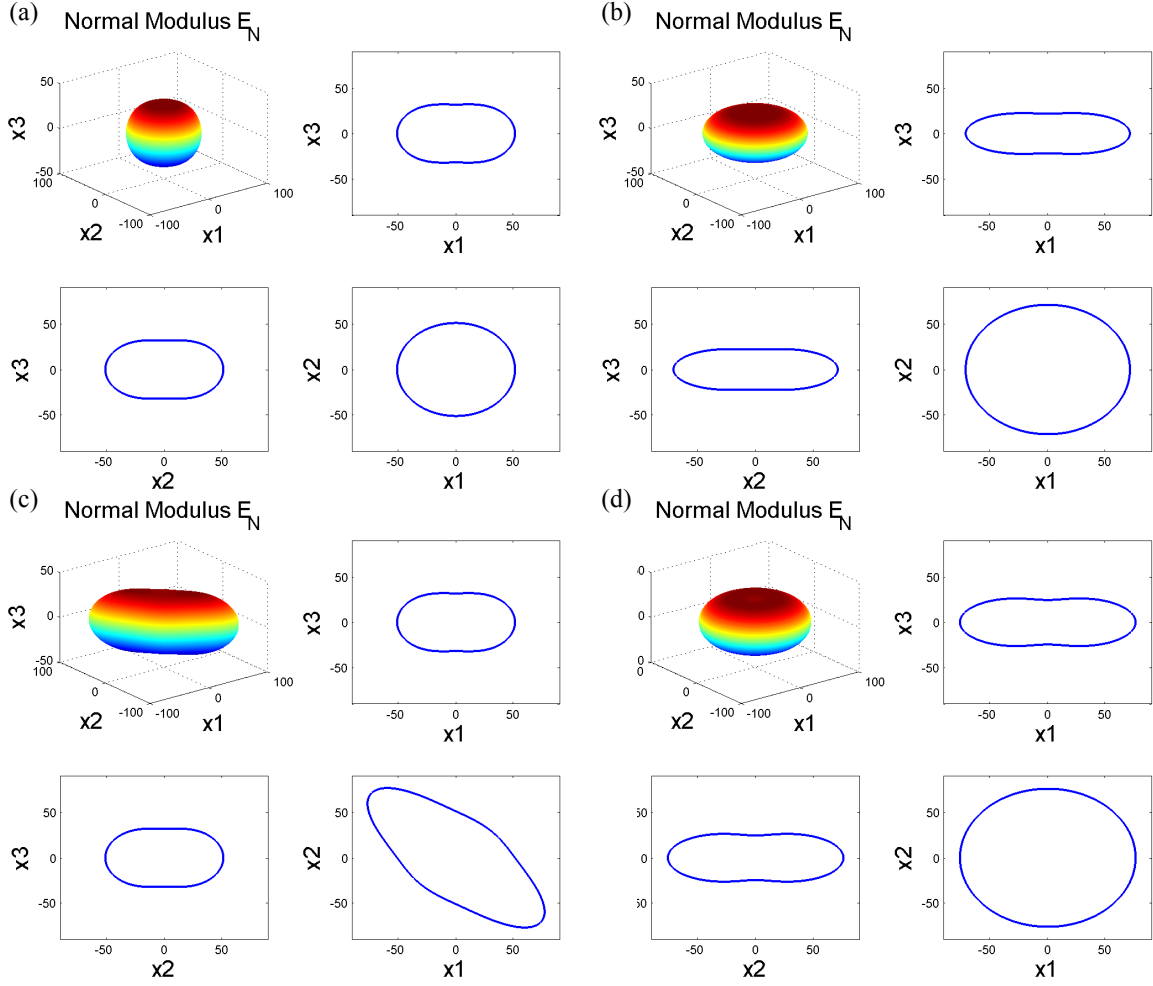}
  \caption{(a) plots the actual $E_N$ under $\epsilon_{11}=\epsilon_{22}=\epsilon_{33}=1$ and $\epsilon_{13}=\epsilon_{23}=\epsilon_{12}=0$; (b) plots the actual $E_N$ under $\epsilon_{11}=\epsilon_{22}=\epsilon_{33}/4=1$ and $\epsilon_{13}=\epsilon_{23}=\epsilon_{12}=0$; (c) plots the actual $E_N$ under $\epsilon_{11}=\epsilon_{22}=\epsilon_{33}=\epsilon_{12}=1$ and $\epsilon_{13}=\epsilon_{23}=0$; and (d) plots $E_N=1/(a_1\sin^2\theta+a_2\cos^2\theta)$ as assumed in Case C. } 
  \label{f10}
  \end{center}
\end{figure}

The actual $E_N$ can be obtained by $E_N=\sigma_N/\epsilon_N$, where $\epsilon_N$ and $\sigma_N$ are given as follows:
\begin{eqnarray}
\label{qqss1}
\epsilon_N=\sin^2\theta[\gamma_1(\cos^2\phi-\sin^2\phi)+2\sqrt{2}\epsilon_6\sin\phi\cos\phi]+\gamma_2(-\sin\omega\sin^2\theta/\sqrt{2}+\cos\omega\cos^2\theta)\\\nonumber+\gamma_3(\cos\omega\sin^2\theta/\sqrt{2}+\sin\omega\cos^2\theta)+2\sqrt{2}\sin\theta\cos\theta(\epsilon_4\sin\phi+\epsilon_5\cos\phi)\\
\label{qqss2}
\sigma_N=\lambda_1\sin^2\theta[\gamma_1(\cos^2\phi-\sin^2\phi)+2\sqrt{2}\epsilon_6\sin\phi\cos\phi]+\lambda_2\gamma_2(-\sin\omega\sin^2\theta/\sqrt{2}+\cos\omega\cos^2\theta)\\\nonumber+\lambda_3\gamma_3(\cos\omega\sin^2\theta/\sqrt{2}+\sin\omega\cos^2\theta)+2\sqrt{2}\lambda_4\sin\theta\cos\theta(\epsilon_4\sin\phi+\epsilon_5\cos\phi)
\end{eqnarray}
with $\gamma_1=(\epsilon_1-\epsilon_2)/2$, $\gamma_2=-\sin\omega(\epsilon_1+\epsilon_2)/\sqrt{2}+\epsilon_3\cos\omega$, and $\gamma_3=\cos\omega(\epsilon_1+\epsilon_2)/\sqrt{2}+\epsilon_3\sin\omega$. Fig.~\ref{f10}(a)-(c) plot the actual $E_N$ under different macroscopic strains, and Fig.~\ref{f10}(d) plots $E_N=1/(a_1\sin^2\theta+a_2\cos^2\theta)$ as assumed in Case C. It can be seen that while the $E_N$ assumed in Case C is a function of $\theta$ only, the actual $E_N$ is a function of not only $\theta$ but also $\phi$ and the macroscopic strain.

\section{Concluding Remarks}
This contribution has studied the extension of the microplane formulation to transversely isotropic materials such as shale rock, foams and ceramics among others. Two possible approaches were investigated, namely: 1) the spectral decomposition of microplane strains; and 2) the introduction of orientation-dependent microplane elastic moduli. 

It was shown that the spectral stiffness decomposition provides the only rigorous approach for the description of microplane strains and stresses in transverse isotropy. However, an approximation almost as accurate could be obtained by making the elastic microplane moduli a function of the microplane orientation. It was shown that the latter approach can span a broad range of macroscopic elastic properties compared to the thermodynamic restrictions on the anisotropic parameters. Further, the approximated functions have the advantage to provide a diagonal microplane elastic matrix which makes easier to guarantee work consistency and, more importantly, makes the formulation of inelastic boundaries easier.

It is concluded that, while the combination of the spectral stiffness decomposition theorem with the microplane approach represents a powerful and rigorous way to capture the elastic behavior of anisotropic materials, it makes the definition of inelastic boundaries slightly more complicated. The spectral stiffness decomposition is recommended in cases such as for unidirectional composites, in which the material behaves almost linearly until failure and an accurate description of the elastic behavior is necessary. Further, it is the only choice when the elastic parameters of the material are beyond the range which can be described by the second approach. For cases in which it is important to capture anisotropy but the material behaves mainly inelastically (as it is the case for shale rock), the second approach is better since it provides a simpler framework for the definition of inelastic boundaries.

\section*{Acknowledgement}
C. Jin thanks the start-up  funds  provided  by  Department of Mechanical Engineering at State University of New York at Binghamton. G. Cusatis thanks the funding support from U.S. National Science Foundation through grant CMMI-1435950 to Northwestern University. 

%
 
\clearpage
\begin{table}[!htp]
\caption{The Five Elastic Constants of Boryeong Shale}
\renewcommand{\arraystretch}{1.0}
\begin{center}
\footnotesize
\begin{tabular}{lccc}
\hline
\hline
Elastic Constants& Experimental Data \cite{3j} & Data Generated by Microplane Model\\
\hline
Young's modulus parallel to bedding, $E$ (GPa)& 34--45.8&37.30\\
Young's modulus perpendicular to bedding, $E'$ (GPa)& 16.5--20.5&18.40\\
Poisson's ratio parallel to bedding, $\nu$ (-)& 0.13--0.23&0.15\\
Poisson's ratio perpendicular to bedding, $\nu'$ (-)& 0.14--0.23&0.16\\
Shear modulus, $G$ (GPa) & 6.2--12.0&12.00\\
\hline
\end{tabular}
\end{center}
\end{table}

\begin{table}[!htp]
	
\caption{The Five Elastic Constants of DIAB Divinycell H60}
\renewcommand{\arraystretch}{1.0}
\begin{center}
	
\footnotesize
\begin{tabular}{p{160pt}>{\centering\arraybackslash}p{160pt}>{\centering\arraybackslash}p{180pt}}
\hline
\hline
Elastic Constants& Experimental Data \cite{foam4,foam5} & Data Generated by Microplane Model\\
\hline
Young's modulus in transverse direction, $E$ (GPa)& 13.0--19.0&16.0\\
Young's modulus in rise direction, $E'$ (GPa)& 31.0--33.0&32.0\\
Poisson's ratio in the plane perpendicular to rise direction, $\nu$ (-)& 0.29--0.31&0.29\\
Poisson's ratio in rise direction, $\nu'$ (-)& 0.04--0.44&0.28\\
Shear modulus, $G$ (GPa) & 15.0--20.0&15.0\\
\hline
\end{tabular}
\end{center}
\end{table}

\clearpage
\clearpage
\clearpage
\clearpage
\clearpage

\clearpage

\clearpage
\clearpage
\clearpage

\clearpage

\clearpage

\end{document}